\documentclass[a4,aps,showpacs,amsmath,floatfix]{revtex4}
\usepackage{graphicx}
\usepackage{dcolumn}
\usepackage{color}
\usepackage{amsmath}
\usepackage{here}
\usepackage{bm}
\usepackage{amsmath}

\begin{document}
\title{Analogues of Josephson junctions and black hole event horizons \\ in atomic Bose-Einstein condensates}
\author{A.I. Yakimenko, O.I. Matsyshyn, A.O. Oliinyk,  V.M. Biloshytskyi, O.G. Chelpanova, S.I. Vilchynskii}
\affiliation{Department of Physics, Taras Shevchenko National University of Kyiv, 64/13, Volodymyrska Street, Kyiv 01601, Ukraine }

%\runninghead{ A. Yakimenko, O. Matsyshyn, A.Oliynyk, V.Biloshytskyi,  O.Chelpanova, S.Vilchynskii}{Analogues of Josephson junctions and black hole event horizons in atomic BEC}

%\maketitle
\begin{abstract}
We study dynamical processes in coherently coupled atomic Bose-Einstein condensates. Josephson effects in ring-shaped and dumbbell geometries are theoretically investigated. Conditions for observation of the Josephson effect are revealed. We found that multicharged persistent current in toroidal condensate can be robust even for supersonic atomic flow. In  numerical simulations  the acoustic analogues of event horizon in quantized superflow was observed.  These theoretical finding open perspectives for investigation of Bose Josephson junctions and quantum aspects of acoustic analogue of Hawking radiation in existing experimental setups.
\end{abstract}


\maketitle

\section{Introduction}

 The phenomenon of superfluidity in  helium liquids $^4$He  and $^3$He,
 as well the phenomenon of superconductivity in the electron Fermi liquid in
metals are fundamental properties of quantum liquids and manifestation of quantum laws
at the macroscopic level.

We have the big progress achieved in the understanding of these phenomena
since the pioneer works by  Bogolyubov \cite{BNN},
Feynman \cite{FR}, Cooper \cite{cooper}  as well  these works \cite{N-P} - \cite{5.30} and now realize that
the essence of the superfluidity and superconductivity phenomena is
the emergence of a certain macroscopic quantity -- the complex order
parameter, which is the wave function of bosons or Cooper pairs of
fermions that occupy the same quantum state.
For a better comprehension on microscopic level
 of  the physical mechanisms of the superfluidity and superconductivity phenomena
we need to understand how these phenomena are
connected with the phenomenon of macroscopic accumulation of bosons
in the ground state of the  Bose-system, which  was
described by Einstein, nowadays known as Bose-Einstein condensation.

The experimental evidence of the existence of Bose-Einstein
condensate (BEC) in the superfluid liquid $^4$He  \cite{sven,Grif,Wyatt}
bolsters one of the main
hypotheses of London and Tisza \cite{London,Tisza},
who tried to show that the superfluidity is closely related
to the motion of the BEC, which
moving as whole.

From this point of view the
experimental realization of the  Bose-Einstein condensation phenomenon
of alkali-metal atoms of rubidium \cite{Anderson} and sodium
\cite{Ketterle} has opened new possibilities for exploring
superfluidity at a much higher level of control.  For the weakly interacting BECs, a relatively simple
Gross-Pitaevskii equation (a variant of nonlinear Schr\"{o}dinger
equation) gives basically good description of the atomic condensates
and their dynamics at low temperatures. It is remarkable that the
strength of interaction can be tuned using Feshbach resonance
\cite{FeshbachRev}, and different geometries of the trapping
potential provide the possibility to study a 2D and even a 1D
system.  A key feature of dilute atomic BEC is that these systems can be efficiently controlled in experiments
by means of laser beams or external magnetic fields. Weak interatomic interactions in dilute gases provides a framework
 for very simple but surprisingly accurate theoretical models.
That is why ultracold gases are actively used as important polygons for emulation of many phenomena in condensed matter physics in a pure form.

  Now atomic BEC is widely used for investigation of a
superfluidity and superconductivity allowing for quantitative tests of microscopic
theories using the tools and precision of atomic physics
experiments. Should be especially  noted that  impressive progress in studies of BECs of ultracold atomic gases gives rise to the new field of research, atomtronics. Electronics deals with the transport and interaction of electrical charges whereas atomtronic devices utilizes neutral atoms. In the future it may be possible to fabricate atomtronic devices analogous to batteries, diodes and transistors, as well as fundamental logic gates. Nowadays, the effects, which are conventional in electronic devices, find their counterparts in BEC. The atomtronic circuit with potential barriers (tunnel junctions or weak links) is suitable for use as an analogue of the superconducting quantum interference device (SQUID), which would sense rotation \cite{2007Natur.449..579L}.

From the other side, many phenomena, previously observed in superfluid and superconductive systems,
have found their counterpart with ultracold
alkali-metal gases.
For instance in
 recent experiments with atomic BEC confined in a dumbbell-shaped potential \cite{2016PhRvA..93f3619E} consisting
of two wells connected by a narrow channel and with toroidal condensate splitted by two weak links \cite{PhysRevLett.113.045305,2013PhRvL.111t5301R}  was found that quantum vortices excited by a super-critical atomic flow leads to effective resistance in this atom circuit.
That opens us the perspective
   to investigate  the possibility of the existence of  analogues of the Josephson effects
(which is  well established and investigated for  different superconductivity systems)
 in ring-shaped and dumbbell geometries BECs and to look for  the  conditions for experimental observation of that effect.
 One of the goal of this work is to  explore this possibility.

Moreover, unique properties of BECs gives us another big possibility to experimental realization of the   way  developed by Unruh
of mapping certain aspects of black hole physics
into problems in the theory of supersonic acoustic flows
(let us remind - black holes  are very hard to experimental research due to  that large astrophysical black holes are too far from us and the small black holes are extremely short-living and haven't been created yet).
In experiment \cite{PhysRevLett.113.045305} optical potential barriers are used to split the atomic cloud into weakly interacting parts. Toroidal condensate with persistent current suggests alternative approach when the atomic flow is separated in subsonic and supersonic regions by acoustic analogues of event horizon. Unruh in his pioneering work \cite{Unruh} has suggested to use the hydrodynamic analogues of black holes for observation of Hawking radiation. Recent experiments \cite{2014,2016} demonstrate creation acoustic analogues of black holes and observation of Hawking radiation in ultracold gases. Another goal of our work is  the
  investigation of quantum aspects of acoustic analogue of Hawking radiation in existing  BEC experimental setups.

Our paper is organized as follows.  In Sec. \ref{SecModel} we describe our model. In Sec. \ref{SecJE} we not only perform 3D numerical simulations of the condensate dynamics for experimental parameters but also compare our results with experiments as well as with previous simu\-lations; second we investigate a possibility of observation of Josephson effects in these systems. In Sec. \ref{SecBH}  we found the conditions for formation of robust supersonic atomic flow in toroidal condensate separated from subsonic region by black hole and white hole event horizons.
%Our findings opened an exciting perspectives for investigation of quantum features of acoustic analogues of black holes

%%%%%%%%%%%%%%%%%%%%%%%%%%%%%%%%%%%
\section{Model}\label{SecModel}
%%%%%%%%%%%%%%%%%%%%%%%%%%%%%%%%%%%
Dynamical properties of ultracold dilute atomic BECs can be accurately described by the mean-field Gross-Pitaevskii  (GP) equation %
%, a variant of well-known nonlinear Schr\"{o}dinger equation:
\begin{equation}\label{GPE}
i \hbar \frac{\partial \tilde\Psi(\textbf{r},t)}{\partial t} = \left[-\frac{\hbar^2}{2M} \Delta + \tilde V(\textbf{r}) + U_0 |\tilde\Psi(\textbf{r},t)|^2 \right]\tilde\Psi(\textbf{r},t),
\end{equation}
where $\Delta$
%$\Delta=\frac{\partial^2}{\partial x^2}+\frac{\partial^2}{\partial  y^2}+\frac{\partial^2}{\partial z^2}$
is a Laplacian operator,
\[
U_0 = \frac{4 \pi \hbar^2 a_s}{M},
 \]
 is coupling strength, $M$ is the mass of the atom, $a_s$ is the $s$-wave scattering length, $\tilde V(\textbf{r})$ is the external trapping potential. The norm of the condensate wave function $\tilde{\Psi}$ is equivalent to the number of atoms:
$
N=\int|\tilde \Psi|^2d\textbf{r}.
$
 Both the number of atoms and energy,
\begin{equation}\label{E3D}
\tilde E=\int\left\{ \frac{\hbar^2}{2M} |\nabla\tilde\Psi|^2  +\tilde V_\textrm{ext}(\textbf{r})|\tilde\Psi|^2+\frac{U_0}{2}|\tilde\Psi|^4\right\}d\textbf{r},
\end{equation}
are the integrals of motion of Eq. (\ref{GPE}) for constant trapping potential $\tilde V_\textrm{ext}(\textbf{r})$.
The steady states of the form $\tilde \Psi(\textbf{r},t)= e^{i\mu t/\hbar}\Psi(\textbf{r})$ obey the stationary GPE,
\begin{equation}\label{GPEstationary}
 \mu  \Psi = \left[-\frac{\hbar^2}{2M} \Delta + \tilde V(\textbf{r}) + U_0 |\Psi|^2 \right]\Psi,
\end{equation}
$\mu$ is the chemical potential. The stationary states in different external traps could be found numerically using imaginary propagation method (ITP).

Following the experimental setups discussed in the next sections we will consider time-dependant trapping potentials when the energy of the system changes with time. We perform numerical simulations of 3D time-dependent GPE with split step fast fourier transform (SFFT) method.

%%%%%%%%%%%%%%%%%%%%%%%%%%%%%%%%%%%
\section{Josephson effects in atomtronic circuits}\label{SecJE}
%%%%%%%%%%%%%%%%%%%%%%%%%%%%%%%%%%
%%%%%%%%%%%%%%%%%%%%%%%%%%%%%%%%%%%%%%%%%%%%%%%%%%%%%%%%%%%%%%
In super\-conductive systems three Josephson effects are well known: direct current (DC) Josephson effect,  alternating current (AC) Josephson effect, and inverse AC Josephson effect (Shapiro effect). The Josephson DC effect is a constant current of atoms through the channel between the wells in the absence of a difference in chemical potentials, owing to tunneling.  In AC Josephson regime at fixed chemical potential difference the phase difference is a linear function of time, thus the flow of atoms will be a sinusoidal signal. In the case if the difference in chemical potentials between the two wells varies periodically, the constant components of the atomic current and the difference of chemical potentials will be discrete quantities at inverse AC Josephson regime.

In this section we will investigate two experiments with ring-shaped BEC separated by two repulsive barriers and dumbbell-shape BEC. Our aim is to find a condition for observation of Josephson effects in these systems.

\subsection{Ring-shaped BEC with two weak links}
 We perform a numerical simulation of the ring-shaped condensate of $^{23}$Na atoms with two moving barriers \cite{PhysRevLett.113.045305}. A tunable chemical potential difference between two weakly coupled condensates is the key feature required for observation of the Josephson effects.   We analyze the mechanism of the transition of the condensate from the superfluid to the resistive regime. It turns out that the vortex excitations appear simultaneously with chemical potential difference for a squeezed in vertical direction potential trap. As the result substantial fluctuations of the phase difference destroys coherence between subsystems of the atomic cloud. It turns out that the vortex excitations do not appear in elongated in vertical direction condensate, which implies the Josephson mechanism for formation of the chemical potential difference.
%---------------------------------
\begin{figure}
\centering
\includegraphics[width=\textwidth]{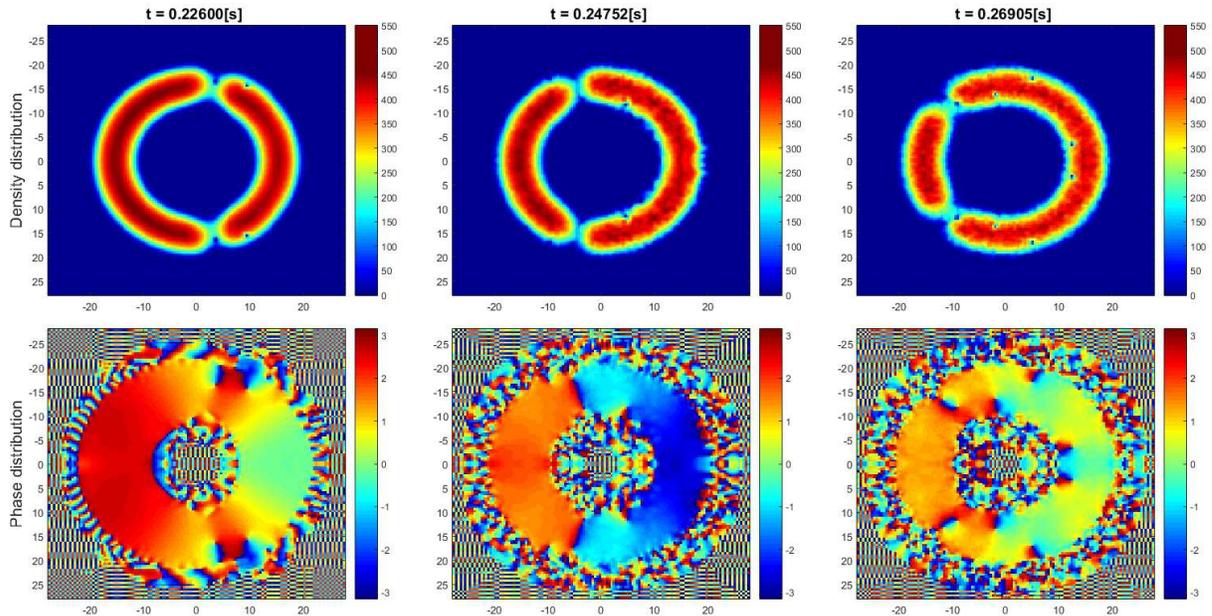}
\caption{Dynamics of the density (top row) and phase (bottom row)  of the ring-shaped condensate with moving weak links for $v=455 \mu$m/s and $z=0$. Few vortex excitations are seen in the subsystem of the ring with lower density.}
\label{fig:image1}
\end{figure}
%-------------------------------
\begin{figure}
\centering
\includegraphics[width=\textwidth]{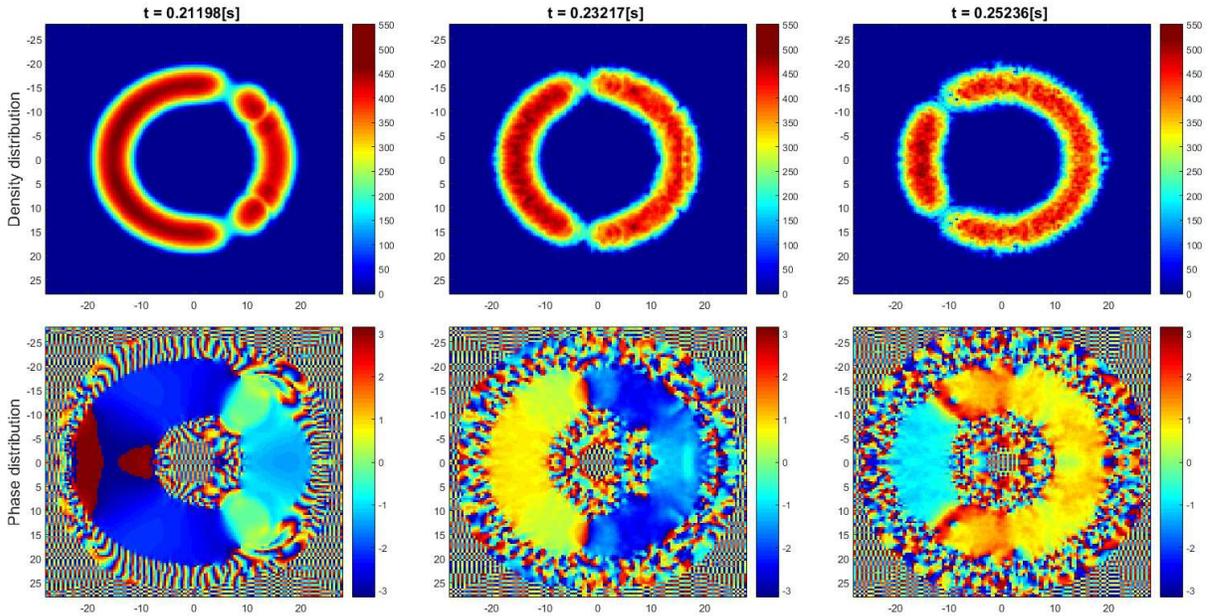}
\caption{The same as in Fig. \ref{fig:image1} for $v=600 \mu$m/s. Note that the number of vortices dramatically increases when the velocity of the barriers grows. }
\label{fig:image2}
\end{figure}
%-------------------------------

The initial condition in  dynamical simulations was taken in the form of stationary state in toroidal trap without barriers. First barrier amplitude rises from zero up to it's maximum value. Barriers arranged symmetrically with angle between them equals to $\pi/2$ .Then barriers begin to move with constant speed in opposite directions (angle between them is growing while evolution). After each barrier scrolled angle equals $\pi/2$ it stops. This procedure will effect on local density of condensate so that the local chemical potential will be changed. Typical examples of the condensate evolution are shown in Fig. \ref{fig:image1} and Fig. \ref{fig:image2} for different values of the velocity of the barriers.

Effectively two barriers splits the system in two weakly interacting subsystems. Assume that each subsystem can be described by local constant chemical potential value, and then calculate chemical potential difference which is a key parameter for Josephson effect. %Here one can dependence of final chemical potential difference on barrier's velocity.
%--------------------
\begin{figure}[h]
\centering
\includegraphics[width=0.75\textwidth]{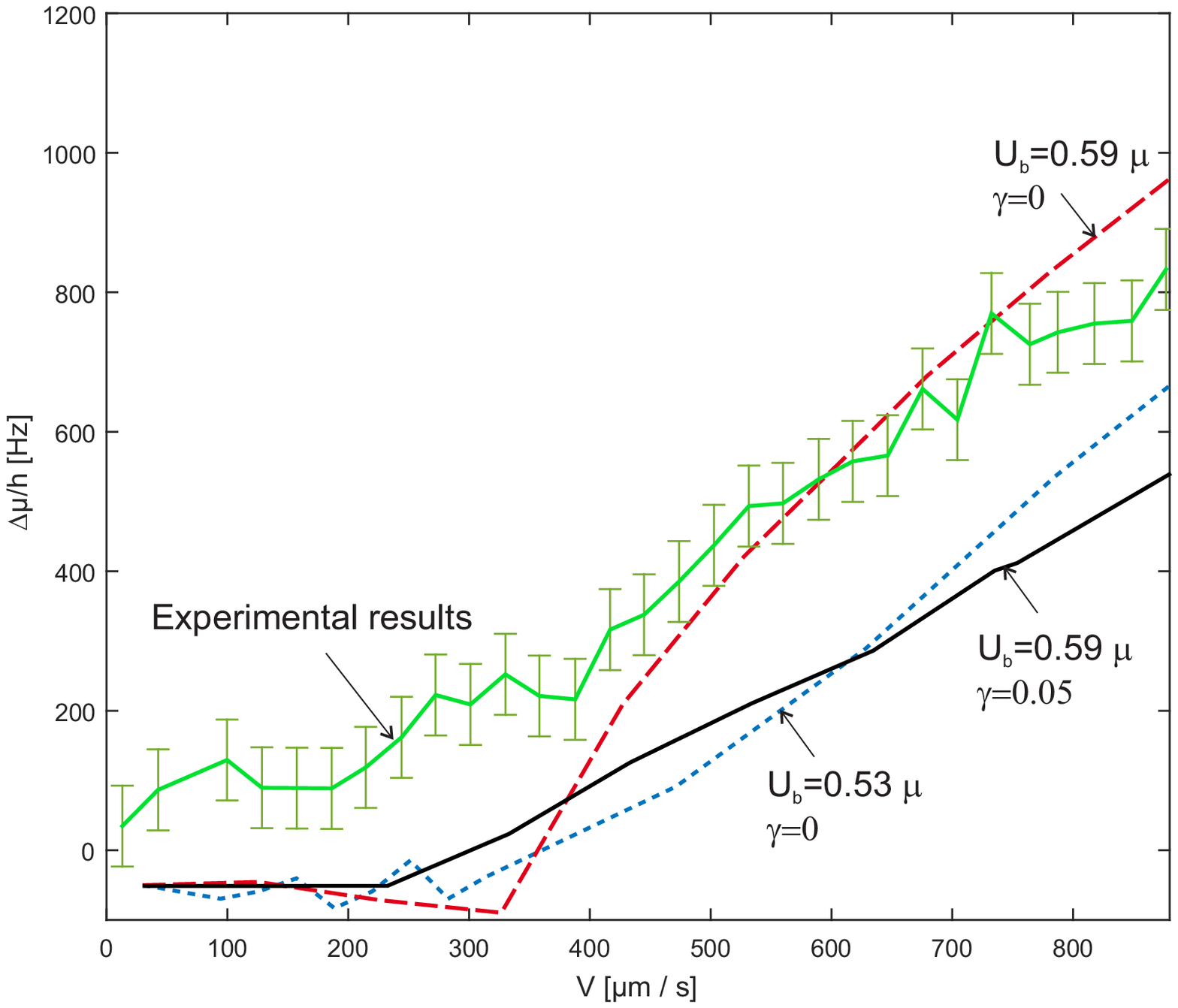}
\caption{Numerically calculated and experimental \cite{PhysRevLett.113.045305}  chemical potential difference  vs velocity of the moving barriers  for different barrier's amplitudes $U_b$ and $\mu/h =  3$ kHz. Note that the chemical potential difference grows rapidly above the threshold value of the velocity $v$ when the vortex excitations appear in dynamics (see Figs. \ref{fig:image1} and \ref{fig:image2} ) }
\label{fig:image3}
\end{figure}
%-------------------------------
To observe Josephson regime it is necessary to create chemical potential difference between coherently coupled systems. Phase correlations rapidly decays if the vortex excitations appear in the atomic flow. Thus we have to work in a regime, when there are no vortices in a system.

Chemical potential difference grows with vortices creation. One can observe vortices on a phase distribution picture above. This regime gives finite chemical potential deference but it is not suitable for Josephson effects observation as a result of subsystems decoherence.
Because of the existence of excited flows, even before the emergence of vortices, there is a viscosity mechanism.
To observe chemical potential difference suitable for Josephson effect in compressed systems one should use lower barrier's velocities and higher barrier's amplitude or width value (to prevent hydrodynamic flow), and in outstretched we have a greater range of options but still we have to ensure that the vortices do not occur.
%-------------------------
\begin{figure}[h]
\centering
\includegraphics[width=0.75\textwidth]{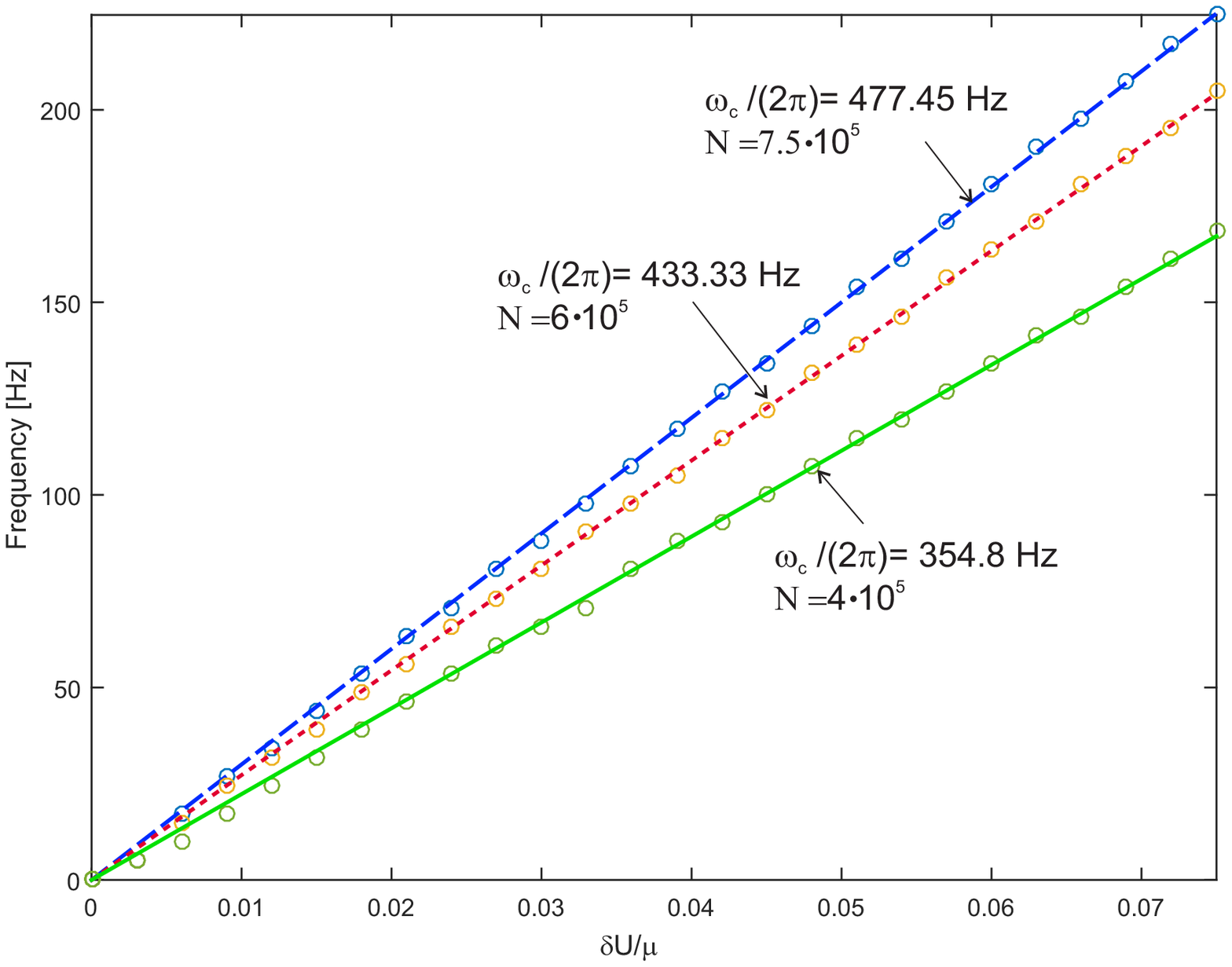}
\caption{Oscillation frequency vs amplitude of the quench (in units of $\mu$) for $U_b=1.68\mu$, $d= 8\xi$, and different values of number of atoms $N$.}
\label{fig:Omega_vs_dU}
\end{figure}
%-------------------------

Here will be discussed another approach with quenching potential. First of all we are creating initial state in asymmetric potential. While the experiment angle between barriers is $\pi$. Barriers amplitude is higher than the value of chemical potential $\mu$.  One consider $N= 1.2\cdot 10^5$ atoms of $^{23}$Na in the toroidal trap with the  following parameters: $R = 20 \mu m,\omega_z = 2\pi \cdot 512 Hz,\omega_r = 2\pi \cdot 260 Hz,U_b = 1.68 \mu,d = 8\xi$, where $\xi$ is the healing length at the peak density.
Two narrow optical barriers separate the toroidal BEC into two weakly coupled subsystems (drain and source) with different amplitude $V_S = V_D + \delta V$, where $ \delta V < 0.1\mu$ to prevent acoustic excitations. Also barrier's potential is modified to make it continuously merging with barrier in another subsystem (barrier's amplitude in different subsystems differs but another parameters are the same).

Next step, after creating initial biased state, we quench potentials to the symmetric subsystems. After that one receive a system in symmetric external potential but with chemical potential difference. One simulate Josephson effect in BEC of $^{23}$Na %. The experiments with  which provides much more control on parameters of the system.
for parameters corresponding to the experiment reported in Ref. \cite{PhysRevLett.113.045305}. %For each of them we observed chemical potential oscillations. %As a result of these simulations one can observe figure:
 As it must be for AC Josephson effect oscillation frequency is a linear function of the chemical potential difference (see Fig.  \ref{fig:Omega_vs_dU}). It is important to note that the studied here condensate with parameters of Ref. \cite{PhysRevLett.113.045305}  with more atoms provides much more control on parameters of the Josephson junction than the small condensate investigated in Ref. \cite{2013PhRvL.111t5301R}.

% Also for these systems and it's oscillation frequencies we can observe such relation (here 1 means for system with 700k particles, 2 for 650k and 3 for 400k):

%\begin{align*}
%    \frac{\mu_1}{\mu_2} = 1.10   &\qquad
%    \frac{ \omega_{c,1}}{\omega_{c,2}} = 1.10\\
%    \frac{\mu_1}{\mu_3} = 1.33    &\qquad
%   \frac{\omega_{c,1}}{\omega_{c,3}}= 1.34
%\end{align*}

%\include{Paper_Rev/intro}
%\include{Paper_Rev/superfl}
%\include{Paper_Rev/microteor}
%\include{Paper_Rev/conv_nonconvBEC}
%\include{Paper_Rev/conclusions}

\subsection{Dumbbell trapping potential}
%%%%%%%%%%%%%%%%%%%%%%%%%%%%%%%%%%%%%%%%%%%%%%%%%%%%%%%%%%%%%%
In this section we consider the condition for appearance of Josephson effects in  optically trapped Bose-Einstein condensate of  $^{23}$Na %\cite{PhysRevA.94.023626}. % One of the objectives of this work is investigation of Josephson effects in atomic BECs.
% It is Shapiro effect.
%----------------------------------------------------
\begin{figure}[h]
\center{\includegraphics[width=\textwidth]{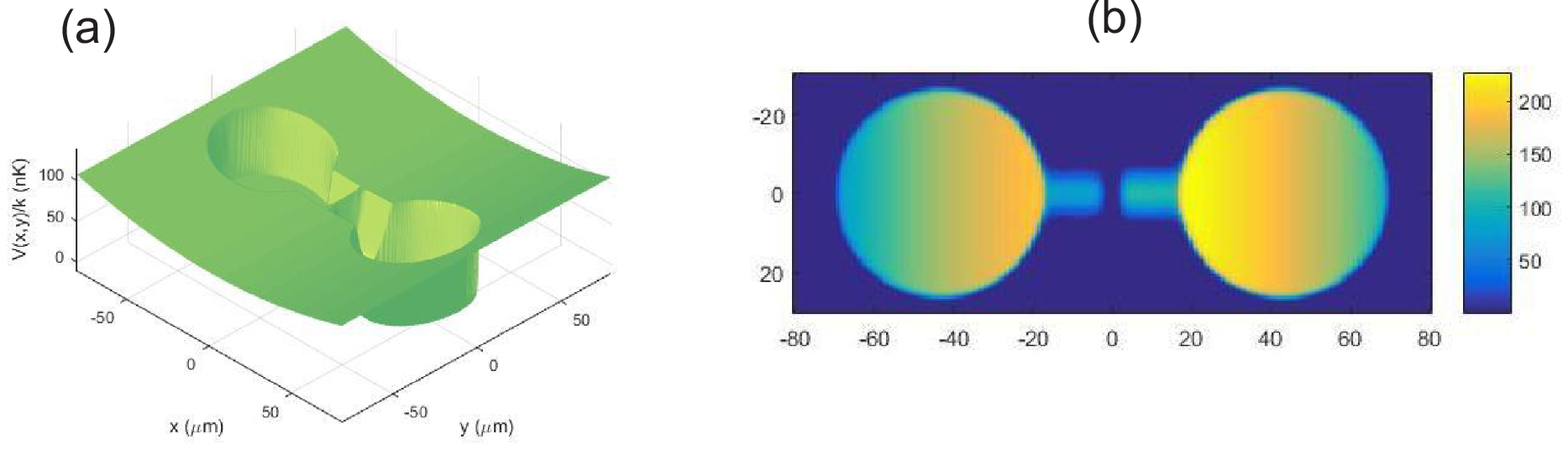}}
\caption{(a) Potential in horizontal $(x,y)$ plane with the optical barrier. (b)  The cross section of the condensate intensity distribution at the plane $z = 0$. The channel between two potential wells is separated by narrow repulsive potential barriers. }\label{Potentials}
\end{figure}
The dumbbell potential $V_{ext}$, which is illustrated in Fig.  \ref{Potentials} (a),
was modeled using parameters reported in Ref. \cite{PhysRevA.94.023626}
\begin{equation}
    V_{ext} = \frac{1}{2}M(\omega_x^2 x^2 + \omega_y ^2 y^2 + \omega_z^2 z^2)+V_{well}+V_{b}+V_{channel}.
\end{equation}

%----------------------------------------------------
To investigate the Josephson effects we use the additional potential  barrier of the form
$$
   V_{b}=U_{a}\exp\left(-\frac{x^{2}}{2a^{2}} \right).
$$
where $U_a/k=90.4$ nK,  $a=1 \mu$m.
%We developed a 3D code to solve this equation. To obtain a stationary solution for this potential we use the numerical method "Imaginary time propagation". To simulate the dynamics of BEC we %use the numerical method "Split-step".

%\newpage
%\section{3-D simulation of strong nonequilibrium state of condensate}

\begin{figure}[h]
\center{\includegraphics[width=\textwidth]{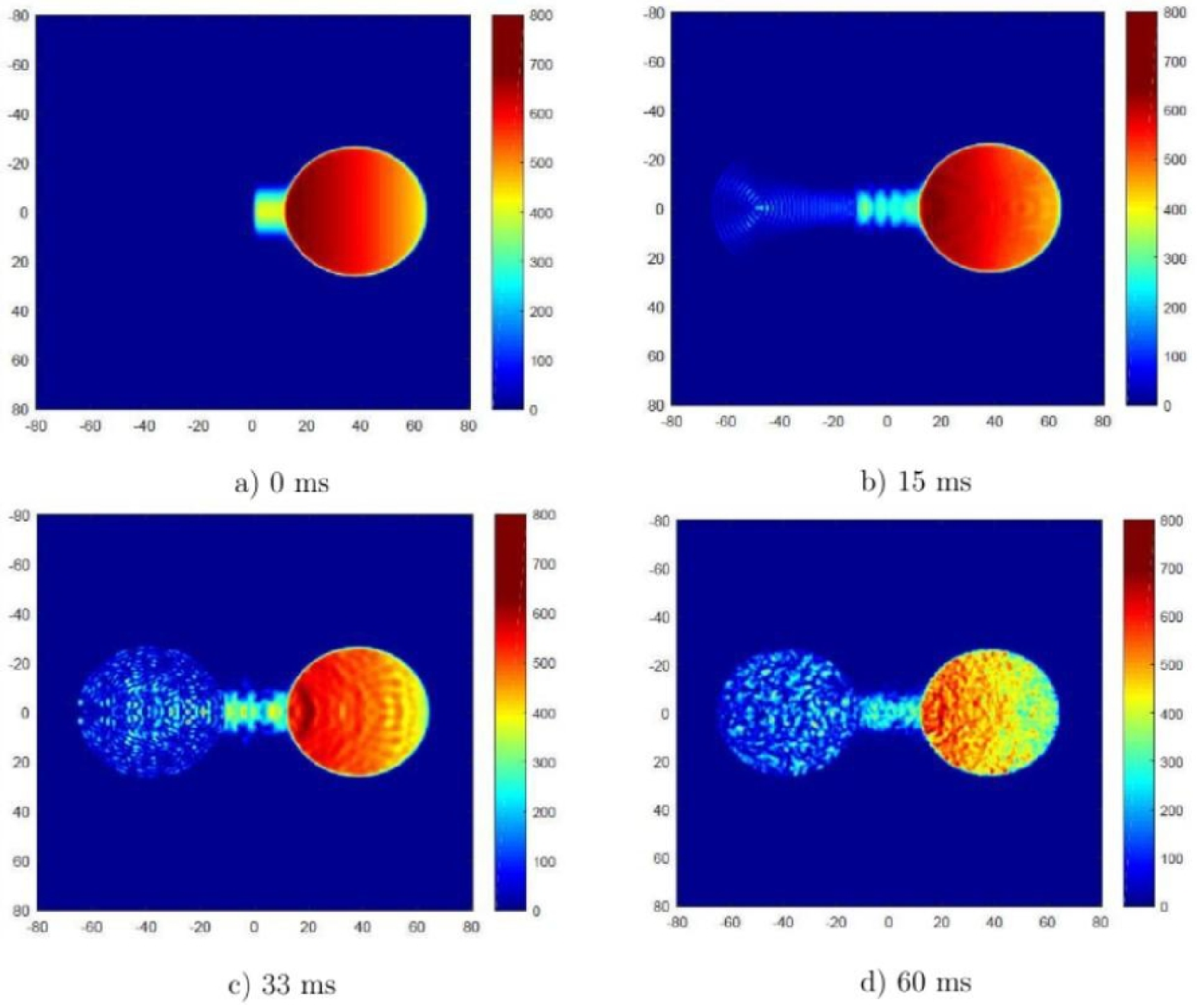}}
\caption{Snapshots of the  condensate density cross sections for $z=0$ for different moments of time. Note the interference patterns for $t= 15 $ms. Finally the system transit in turbulent state when the supercritical flow generates many vortices.} \label{dynamicsDumbbel}
\end{figure}

If we put $N$ particles in the right well and no particles in left-hand side part (see Fig. \ref{dynamicsDumbbel}) the flow appears from the right well to the left since the number of atoms in different wells tends to equalize.
%-------------------------------------
%--------------------
%----------------------------
\begin{figure}[h]
\center{
\includegraphics[width=\textwidth]{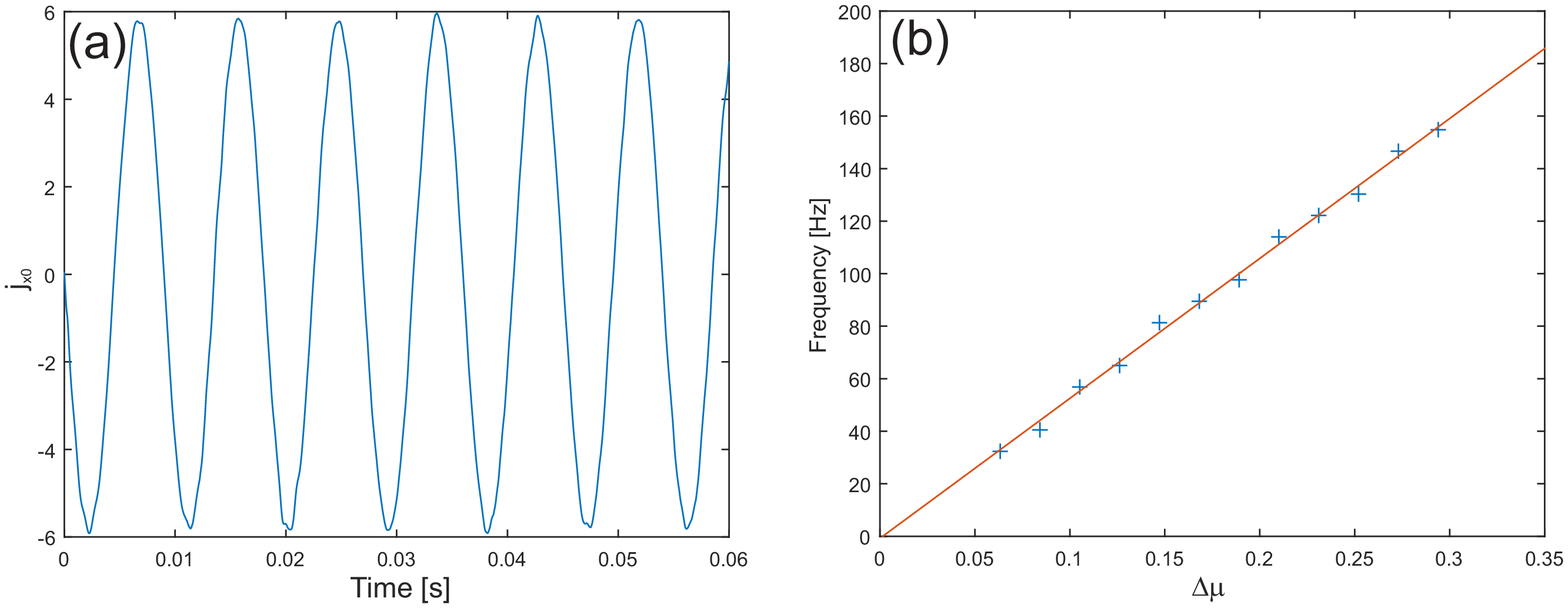}}
\caption{Numerical results for AC Josephson effect in the dumbbell-shape BEC. (a) Superflow current through potential barrier vs time. (b) Frequency of AC vs dimensionless chemical potential difference $\delta\mu/(\hbar\omega_z$).}\label{fig:ACJE}
\end{figure}
%----------------------------

We have performed 3D simulations from the parameters of Ref. \cite{PhysRevA.94.023626} and compare our results with 2D modeling  reported in this section. Our results are found to be in good agreement with previous 2D simulations. But when there is movement of condensate along the $z$ axis or compression of the condensate on the $z$ axis is not big, then 2D code does not always work well. So, our result is more precise, and the method is more general. The interference patterns and vortices  of the condensate in turbulent regime are clearly seen in Fig.  \ref{dynamicsDumbbel}. It is strong nonequilibrium state of condensate, because of the large initial difference in the number of atoms in different wells.

%\newpage
%\section{AC Josephson effect in the dumbbell potential}
As was mentioned above, for observation of  the Josephson regime one need to create a nonzero chemical potential difference between two wells. One put a different number of atoms in different wells. Method for creating an imbalance of the number of atoms is as follows. First one find a stationary solution with biased potential (add to the potential of the left well, for example, a constant value $U_{delta}$), after that one simulate the dynamics with the unbiased potential. Typical example with the results of the numerical simulations is illustrated in Fig.\ref{dynamicsDumbbel}. For each value $U_{delta}$ in this method one can calculate corresponding chemical potential difference using Thomas-Fermi approximation.

If one creates the initial state with $N_{1}$ atoms in the left well and $N_{2}$ atoms ($N_{2}\neq N_{1}$) in the right well, then the periodic signal for difference between number of atoms in left and right wells is observed. In  Fig. \ref{fig:ACJE} (a) one illustrate a typical dynamics of the superflow component $j_{x0}$, where
$\textbf{j}=-\frac{i}{2}\int\left(\Psi^*\nabla\Psi-\Psi\nabla\Psi^*\right)d\textbf{r}$.  To confirm that these oscillations correspond to AC Josephson effect, we have performed an extensive series of numerical simulations for different values $U_{delta}$ (which corresponds to various chemical potential difference). During all simulation time the difference in chemical potentials remains approximately constant since we have used small values of potential bias $U_{delta}\ll V_{well}$. It is important to note that in this regime no  turbulent flow appears. The frequency of oscillations in the atomic flow was extracted from the periodical signal  with Fast Fourier Transform (FFT). The frequency as the function chemical potential difference is shown in Fig. \ref{fig:ACJE} (b). Solid line is mean-square approximation of the numerical data (crosses). The numerical estimate gives for the frequency  $\omega/(2\pi)=532.9$ Hz, which is an excellent agreement with theoretically predicted value  $\omega/(2\pi) =$ 529Hz. %This is a good result: we modeled the AC Josephson effect.

%\newpage
%\section{Conclusion}

%%%%%%%%%%%%%%%%%%%%%%%%%%%%%%%%%%%%%%%%%%%%%%%%%%%%%%%%%%%%%%
\section{Acoustic analogue of event horizon in toroidal BEC}\label{SecBH}
%%%%%%%%%%%%%%%%%%%%%%%%%%%%%%%%%%%%%%%%%%%%%%%%%%%%%%%%%%%%%%
Hawking radiation is one of the most astonishing and hard-to-research effect in our macroworld. Large astrophysical black holes are too far from us, while the small black holes, which could be observed in the high-energy experiments, are extremely short-living and haven't been created yet. Thus in according to W. Unruh \cite{Unruh} a creation and investigation the analogues of black holes is only one way to investigate black hole effects experimentally for today. In this section we simulate 3D toroidal BEC, where stable closed flow with supersonic region would be created easily.

Taking into account recent theoretical investigations \cite{2007,2001} of ring-shaped BEC systems with 1D supersonic flow we have found optimal way to create a stable 3D flow in toroidal geometry.
In  simulations we use the parameters which are typical for recent experiments with BEC: $N = 2 \cdot 10^4$ $^{87}$Rb atoms with mass $M = 1.45\times 10^{-25}$kg and the s-wave scattering length $a = 5.77\times 10^{-9}$m \cite{exper}.

The first step of investigation contains finding the stable time-dependent solution of GP \eqref{GPE}
with external adiabatically time-dependent potential
\begin{equation}\label{generalized p}
V(\textbf{r};t) = V(r,\varphi,z;t) = V_{trap}(r,z) + V_{nozzle}(\varphi;t),
\end{equation}
where
\begin{equation}\label{trap p}
V_{trap}(r,z) = \frac{1}{2}M\omega^{2}_{z} z^{2}  + \frac{1}{2}M\omega^{2}_{r}\left( r - R \right)^{2},
\end{equation}
\begin{equation}\label{nozzle p}
V_{nozzle}(\varphi;t) = -V_{0}\left(t+(1-t)\theta(t-t_{0})\right)\cos{\varphi}.
\end{equation}
Here $R$ is the radius of the toroidal trap, $t_0$ corresponds to time of growing of the amplitude of the  potential $V_{nozzle}$.

The trap potential $V_{trap}(r,z)$ is the stationary oscillator potential along the $z$- and $r$-directions with corresponding frequencies $\omega_{r}$ and $\omega_{z}$.
This is the potential of unperturbed trap, where the subsonic condensate flow will exist.
$V_{nozzle}(\varphi;t)$ considered as a small addition to the $V_{trap}(r,z)$ , is the angle- and time-dependent potential for nozzle creation and consequently supersonic flow production.

Defining overall solution as
\begin{equation}\label{solution form}
\psi(\textbf{r},t) = \widetilde{\Psi}(\textbf{r},t)\exp(-i\widetilde{\mu} t),
\end{equation}
%and substituting it into \eqref{GPE} yields
%\begin{equation}\label {gp mu}
%i\hbar\partial_{t}\widetilde{\Psi}(\vec{r},t) = \left(-\frac{\hbar^{2}}{2m}\nabla^{2} + V(\vec r;t) + g|\widetilde{\Psi}(\vec{r},t)|^{2} - \widetilde{\mu}(t)\right)\widetilde{\Psi}(\vec{r},t),
%\end{equation}
and  GPE is rewrited in terms of harmonic oscillator units: $t\rightarrow\omega_{r}t,\textbf{r}\rightarrow\textbf{r}/l_r, V \rightarrow  V/(\hbar\omega_{r}), g \rightarrow g/(\hbar\omega_{r} l_{r}^3),\Psi = \widetilde{\Psi}\cdot l_{r}^{\frac32},\mu = \widetilde{\mu}/(\hbar\omega_{r})$, where $ l_{r} = \sqrt{\hbar/M \omega_{r}}$, we have
%--------
\begin{equation}\label {gp mu dimless}
i\partial_{t}\Psi(\textbf{r},t) = \left(-\frac12\nabla^{2} + V(\textbf r;t) + g|\Psi(\textbf{r},t)|^{2} - \mu(t)\right)\Psi(\textbf{r},t).
\end{equation}
%--------
The parts $V_{trap}$ and $V_{nozzle}$ of the potential term $V$ are acquiring shape
\begin{eqnarray}\nonumber\label{dimless trap nozzle p}
V_{trap}(r,z) = \frac{1}{2}\kappa^{2} z^{2}  + \frac{1}{2}\left( r - R \right)^{2},
\\
V_{nozzle}(\varphi;t) = -V_{0}\left(t+(1-t)\theta(t-t_{0})\right)\cos{\varphi},
\end{eqnarray}
from the \eqref{trap p} and \eqref{nozzle p} with the following replacement and substitution: $V_{0} \rightarrow  V_{0}/(\hbar\omega_{r})$,  $R\rightarrow R/l_{r}$, $\kappa = \omega_{z}^2/\omega_{r}^2$, where $\theta(x)$ is the Heaviside step function.
The main parameters of  model and their values are as follows:  $R = 4\, \mu$m, $\omega_{r}=2\pi\times 800\,$Hz, $\omega_{z}$ = $2\pi\times400$ Hz.

%summarized in the Table \ref{table:parameters}.

%\begin {table}[h!]
%\centering

%\begin{tabular}{|c|c|}
	%\hline
%	Model parameter & in dimensionless form  (in SI) \\
	%\hline
%	$R$ & $10.5158$  ($4\,\mu m$) \\
%	\hline
%	$g$ & $0.1906$ ($5.5612\times10^{-51}\,J\cdot m^3$) \\
%	\hline
%	$\omega_{r}$ & $2\pi\times800$ ($2\pi\times800\,Hz$) \\
%	\hline
	%$\omega_{z}$ & $2\pi\times400$ ($2\pi\times400\,Hz$) \\
	%\hline
	%$V_{0}$ & $5$ \\
	%\hline
	%$N$ & $20000$ \\
	%\hline
%\end{tabular}

%\caption {Numerical values of the model parameters in dimensionless form and in SI}
%\label{table:parameters}

%\end{table}

%\section{Simulation}

\begin{figure}[h]
	\centering
	\includegraphics[width=\textwidth]{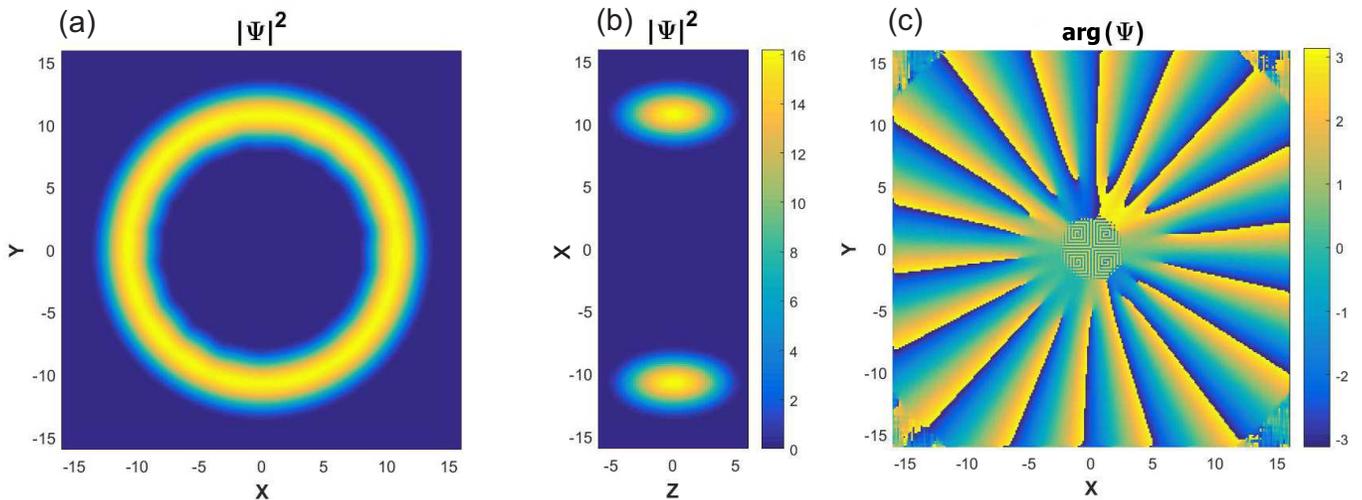}
	\caption{Density $|\Psi(x,y,z=0)|^2$, $|\Psi(x = 0,y,z)|^2$ and phase distribution $arg(\Psi(x,y,z=0))$ of the stationary solution of the \eqref{gp mu dimless} without $V_{nozzle}$, found numerically for $\mu = 4.62$. Persistent current in this case corresponds to topological charge $m=18$. }\label{density phase itp}
\end{figure}

Creation of the supersonic region in the persistent flow was implemented in the following order. First of all we found the stationary solution of the \eqref{gp mu dimless} without $V_{nozzle}$ using ITP-method. The vortex phase was imprinted in the initial wave function $\Psi$
\begin{equation}\label{initial psi}
\Psi(\textbf{r},0) = \Psi(r,z,\varphi,0) =  \Phi(r,z,\varphi)\cdot e^{im\varphi}
\end{equation}
with vortex charge $m = 18$, where $\Phi(r,z,\varphi)$ was the 3D complex function with random distributed amplitude and phase. The choice of charge value follows from condition of the sonic flow, because we want to move our initial solution closer to the solution with the sonic flow.   Results are represented in Fig.\ref{density phase itp}.

%--------------------------
\begin{figure}[h]
	\center\includegraphics[width=\textwidth]{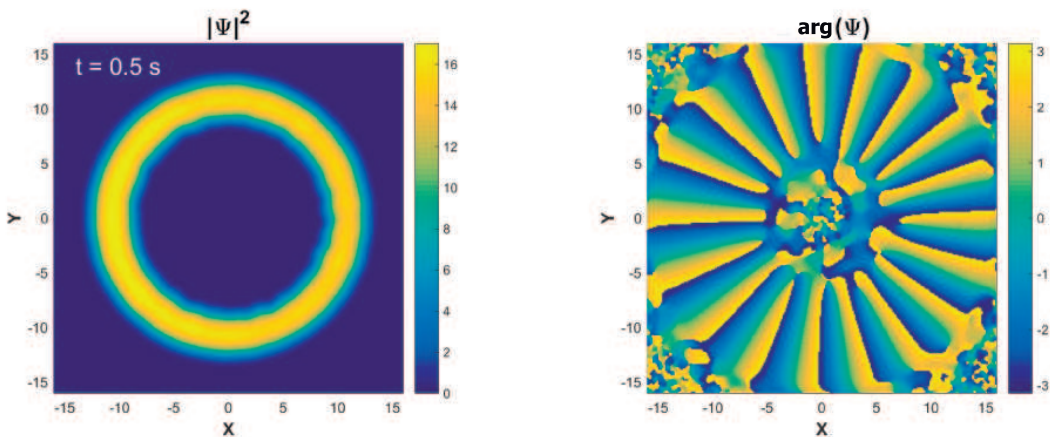}
	\caption{Density $ |\Psi(x,y,z=0)|^2$ (left) and phase distribution $arg(\Psi(x,y,z=0))$ (right) of the perturbed flow after the long-term evolution (for $t=0.5$ s). }\label{0.5}
\end{figure}
To estimate the magnitude of the flow inside the trap  we  found  the value of the flow velocity in the bottom of the trap and compared it with the local sound velocity. It was found that the value of Mach number for the considered configuration of the flow is approximately 0.975.
%-----------------------------
\begin{figure}[h!]
	\center\includegraphics[width=\textwidth]{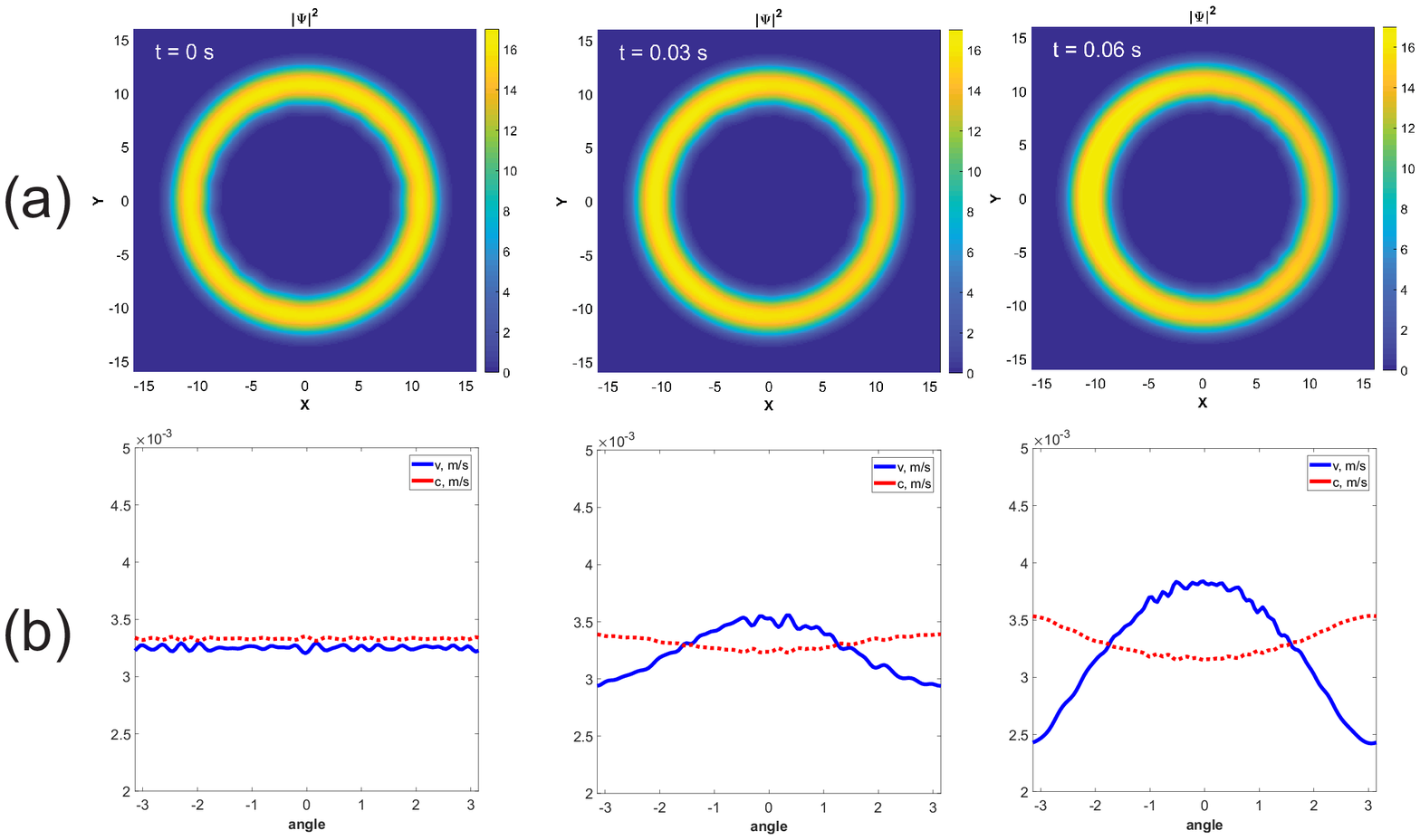}
	\caption{Snapshots of the superflow dynamics for different moments of time: (a)  Density profiles $|\Psi(x,y,z = 0)|^2$. (b)  Flow velocity $v$ and the speed of sound $c_s$ at the bottom of the toroidal trap ($r=R$).  During the time evolution with monotonically increasing potential $V_{nozzle}$ ($t_{0}$ is equals to duration of simulation. The black hole and white hole horizons appears when $v=c_s$.)}\label{rho v c on time}
\end{figure}

%\begin{figure}
%	\centering
	%\includegraphics[width=0.4\textwidth]{vcITP}
	%\caption{Flow and sound velocity distribution in the bottom of the toroidal trap depending on the angle.}\label{v c itp}
%\end{figure}

%
%\begin{figure}[h!]
%\center\includegraphics[width=8.5cm]{vcITP}
%  \caption{Flow and sound velocity distribution in the bottom of the toroidal trap depending on the angle}\label{v c itp}
%\end{figure}

Having found the required stationary solution, we assessed its stability by propagation in time using SSFT-method during 1 sec, as a quite long time interval for real BEC systems.
Should be noted that all the integrals of motion (the number of atoms $N$, the Hamiltonian $H$, the $z$-projection of the angular momentum per particle (or the charge of vortices $L_{z}/N$)) conserved well.

%\begin{figure}[h!]
	%\centering
	%\includegraphics[width=0.32\textwidth]{Nt}
	%\includegraphics[width=0.32\textwidth]{Lt}
	%\includegraphics[width=0.32\textwidth]{Ht}
	%\caption{Proof of the stationary solution stability. Behavior of the general system integrals of motion $N$, $L_z/N$, $H$ depending on time.}\label{N L H}
%\end{figure}

Then we added a perturbation therm $V_{nozzle}$ to the total unperturbed potential $V$ and discovered its influence at the time stability of the founded solution.

%\begin{figure}
	%\centering
	%\includegraphics[width=0.4\textwidth]{Graph1}
	%\caption{Charge number $m = L_z/N$ vs. time}\label{charge on time}
%\end{figure}

Amplitude $V_{0}$ of the potential addition $V_{nozzle}$ was constant. Parameter $t_{0}$ in Eq. \eqref{dimless trap nozzle p} had the same value throughout the period of simulation.
It is seen from Fig.\ref{rho v c on time}  that the borderline value of the time $t$, when the flow is still stable, is up to 40 sec, so we chose $t_{0} = 30$ sec for the further simulations. It allowed us to work with a stable supersonic region, which stability is proved by Fig.\ref{0.5}. It turns out that systems with stretched on $z$-direction and radially compressed toroidal trap potential are more stable than others and allowed to create multi-charged flows with quite high topological charge. It turns out that minimizing radius of the trap also stabilizes the acoustic horizons.

%\newpage
%%%%%%%%%%%%%%%%%%%%%%%%%
\section{Conclusions}
%%%%%%%%%%%%%%%%%%%%%%%%%
%
The different dynamical coherent effects in weakly  coupled Bose-Einstein condensates were  studied -
we consider atomic clouds in toroidal and single-connected (dumbbell) optical traps. It were demonstrated that ring-shaped condensate with moving barriers and supercritical atomic flow in dumbbell condensate exhibits vortex excitations which destroys coherent coupling between coupled subsystems.
 %We suggest a method for observation of  Josephson effects in atomic BEC under realistic experimental conditions.
  Using rapid quench of the trapping potential, we prepare constant chemical potential difference and observe oscillating atomic flow through optical barrier. We have performed extensive series of numerical simulations for different number of atoms and different quench amplitudes. The spectrum of the periodical current observed in  simulations corresponds to AC Josephson regime.  These theoretical results not only clarify the physical background for the experiments with Josephson effects in atomic BEC, but also suggest the parameters of the condensate which can be used for observation of the AC Josephson effect in accessible atomtronic circuits.

 Also we have studied the  strong nonequilibrium dynamics of  optically trapped Bose-Einstein condensate of  $^{23}$Na atoms in dumbbell potential. It turns out that results of  3D simulations of the experimental observations are in good agreement not only with experiments but also with approximate 2D simulations known in literature. Using numerical modelling we demonstrate that  AC Josephson effect can be observed in such atomic BEC in dumbbell potential.
 %Results: the graph of the dependence of the oscillation frequency on the difference in chemical potentials showed that we have a real AC Josephson effect.
 It was shown that oscillation frequency of the superfluid through a narrow potential barrier is proportional to the chemical potential difference between the sub-systems of the dumbbell BEC. This observation clearly demonstrate existence of AC Josephson effect in the system under consideration.

 We create in ring-shaped condensate dynamical event horizons separating the multicharged persistent current into two regions with subsonic and supersonic atomic flows. We revealed realistic parameters of the toroidal potential where partially-supersonic stable flows with well-defined horizons exist.
 These results provide evidence that long-lived acoustic horizons in toroidal BEC can be created with state-of-the-art technology. There are numerous directions along which it would be interesting to continue the present study. For example it of fundamental interest to investigate Hawking radiation in a quantized persistent current. These studies are under present consideration and will be reported in future presentations.

In conclude we note that results, which were obtained in this work,
   open new perspectives for experimental and theoretical investigation of the Bose Josephson junctions and acoustic analogues of black holes.

%%%%%%%%%%%%%%%%%%%%%%
\section*{ACKNOWLEDGMENTS}
%%%%%%%%%%%%%%%%%%%%%%%%%
The authors are grateful to A. Desyatnikov and V. Cheianov  for discussions and comments about this paper.
V.S. is grateful for the support of this work to the German Academic Exchange Service (DAAD), Grant No. 91563279 (Scholarship Programme
"Research Stays for University Academics and Scientists, 2015"). Y.A., B.V. and O.A. acknowledge support from Project $1/30-2015$ 'Dynamics
and topological structures in Bose-Einstein condensates of ultracold gases' of the KNU Branch Target Training at the NAS of Ukraine.

%\bibliographystyle{unsrt}

%\bibliography{refsFNT2018}

\end{document}